\documentclass[11pt]{article}
\usepackage{graphicx}
\usepackage[margin=1.25in]{geometry}
\usepackage[usenames,dvipsnames]{color}
\usepackage{url}
\usepackage[colorlinks = true,
            linkcolor = blue,
            urlcolor  = blue,
            citecolor = blue,
            anchorcolor = blue]{hyperref}

\newcommand\pubnumber{TBD}
\newcommand\pubdate{\today}

\def\Title#1{\begin{center} {\LARGE #1 } \end{center}}
\def\Author#1{\begin{center}{ \sc #1} \end{center}}
\def\Address#1{\begin{center}{ \it #1} \end{center}}

\newcommand\pubblock{\rightline{\begin{tabular}{l} \pubnumber\\
         \pubdate \end{tabular}}}
\newenvironment{Abstract}{\begin{quotation} \begin{center}
                       ABSTRACT
     \end{center}\bigskip  }{\end{quotation}}

\def\beq{\begin{equation}}
\def\eeq#1{\label{#1}\end{equation}}
\def\eeqn{\end{equation}}

\newenvironment{Eqnarray}%
   {\arraycolsep 0.14em\begin{eqnarray}}{\end{eqnarray}}
\def\beqa{\begin{Eqnarray}}
\def\eeqa#1{\label{#1}\end{Eqnarray}}
\def\eeqan{\end{Eqnarray}}

\let\bar=\overbar

\def\lsim{\mathrel{\raise.3ex\hbox{$<$\kern-.75em\lower1ex\hbox{$\sim$}}}}
\def\gsim{\mathrel{\raise.3ex\hbox{$>$\kern-.75em\lower1ex\hbox{$\sim$}}}}

\def\del{\partial}
\def\Dslash{\not{\hbox{\kern-4pt $D$}}}
\def\dslash{\not{\hbox{\kern-2pt $\del$}}}
\def\pslash{\not{\hbox{\kern-2pt $p$}}}
\def\ETmiss{\not{\hbox{\kern-4pt $E$}}_T}

\def\Dlr{\mathrel{\raise1.5ex\hbox{$\leftrightarrow$\kern-1em\lower1.5ex\hbox{$D$}}}}

\def\MSB{{\bar{M \kern -2pt S}}}
\def\msb{{\bar{\scriptsize M \kern -1pt S}}}

\def\drb{{\bar{\scriptsize D \kern -1pt R}}}

\newcommand\snowmass{\begin{center}\rule[-0.2in]{\hsize}{0.01in}\\\rule{\hsize}{0.01in}\\
\vskip 0.1in Submitted to the  Proceedings of the US Community Study\\ 
on the Future of Particle Physics (Snowmass 2021)\\ 
\rule{\hsize}{0.01in}\\\rule[+0.2in]{\hsize}{0.01in} \end{center}}

\begin{document}

\pubblock

\Title{The SLAC Linac to ESA (LESA) Beamline for Dark Sector Searches and Test Beams}

\bigskip 

\Author{Tom Markiewciz, Tor Raubenheimer\footnote{torr@stanford.edu}, Natalia Toro,\\ and members of the LESA construction team}

\medskip

\Address{SLAC National Accelerator Laboratory, Menlo Park, CA}

\medskip

 \begin{Abstract}
\noindent The Linac to End Station A (LESA) beamline is being constructed at SLAC and will provide a near-CW beam of multi-GeV electrons to the SLAC End Station A for experiments in particle physics. The 1st half of LESA is ready for commissioning in FY23 and the full beamline will be operational in FY24. The low-current multi-GeV electron beam is produced parasitically by the superconducting RF (SRF) linac for the LCLS-II/LCLS-II-HE X-ray Free Electron Laser.  LESA is designed to host experiments to detect light dark matter such as the Light Dark Matter eXperiment (LDMX) as well as a wide range of other experiments and test beams requiring near-CW electron currents ranging from pA to $\mu$A. 

\end{Abstract}

\snowmass

\def\thefootnote{\fnsymbol{footnote}}
\setcounter{footnote}{0}

\section{INTRODUCTION}
The SLAC Linac to End Station A (LESA) beamline is a staged concept to provide a near-CW beam at 186 MHz and sub-harmonics thereof to the SLAC End Station A (ESA) for experiments in particle physics requiring pA to 25 nA electron beams with multi-GeV electron energy.  This capability is achieved parasitically by extracting unused bunches from the LCLS-II/LCLS-II-HE superconducting RF (SRF) linac.  The LESA beamline has also been known as DASEL and S30XL and the beamline is described in detail in \cite{Raubenheimer:2018mwt} and summarized in the Snowmass LOI \cite{S30XL_AF5_LOI}.

The LCLS-II is an X-ray Free Electron Laser (FEL) that has been constructed and is being commissioned at SLAC during FY23.  It is based on a 4.0 GeV 1.3 GHz SRF linac \cite{LCLSIIFDR}, fed by an RF gun \cite{LCLSIIgun} operating at 186 MHz. The LCLS-II SRF linac will be upgraded in FY27 to 8 GeV by the LCLS-II-HE project \cite{LCLSIIHECDR}\cite{LCLSIIHEDeltaCDR} which is under construction. 

The baseline LCLS-II/LCLS-II-HE design has a maximum bunch rate of 929 kHz where the bunches are spaced by 1,400 1.3 GHz linac RF buckets. These bunches are diverted from the dump line to the Hard and Soft X-ray (HXR and SXR) undulators by high-speed kickers. LESA will extract unused bunches out of the LCLS-II beamline through the SLAC Beam Switch Yard (BSY) and into the A-line to transport them to ESA.  The layout of the LESA extraction is shown in Figure 1. 

\begin{figure}
\begin{center}
\includegraphics[width=0.70\hsize]{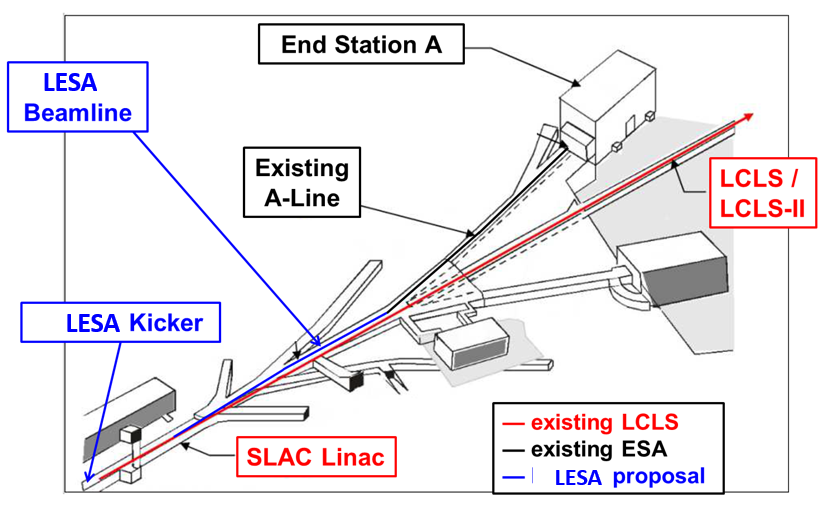}
\end{center}
\caption{Layout of LCLS-II and LESA at the end of the SLAC linac and in the BSY and ESA.}
\label{fig:Fig1}
\end{figure}

\begin{figure}
\begin{center}
\includegraphics[width=0.95\hsize]{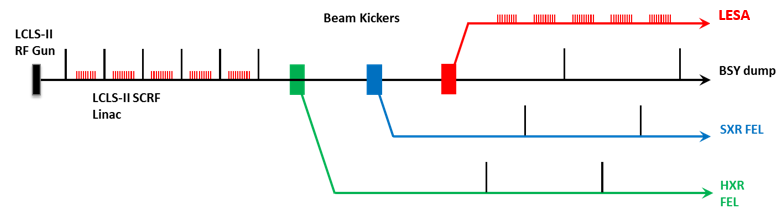}
\end{center}
\caption{Schematic of the LESA beam extraction from the LCLS-II superconducting linac. The LESA beamline directs unused beam to End Station A downstream of the extractions to the LCLS-II HXR and SXR undulators to avoid interference with the primary photon science program.}
\label{fig:Fig2}
\end{figure}

\begin{figure}
\begin{center}
\includegraphics[width=0.95\hsize]{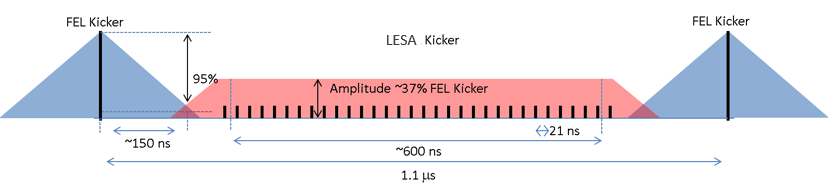}
\end{center}
\caption{LCLS-II pulse structure showing primary pulses with $~10^8$ e- and LESA bunches from the gun with ~30 e- per bunch. The LESA beam will control the bunch population with an additional seed laser and/or a spoiler/collimation system to deliver final current in the pA to $\mu$A range.}
\label{fig:Fig3}
\end{figure}

The LCLS-II FEL bunches are supplied by an RF gun operating at 1/7 of the linac RF frequency which is designed to provide an average current of 30$\,\mu$A. 
Even at full-rate operation, the FEL program uses a tiny fraction of the RF gun and linac repetition rate.  In between every two LCLS-II FEL bunches there are 200 ``empty'' RF buckets from the gun. These empty RF buckets are populated by dark current originating from the gun.  The accelerated dark current has not yet been measured but is expected to be at the level of 10 nA.  In addition, the existing gun laser oscillator could be used to produce a well-defined, low-current beam at 46 MHz repetition rate within a $\sim$500 ns macro-pulse between the primary bunches spaced at 1.1 $\mu$s.  The extraction concept is illustrated in Figures 2 and 3.

The LESA beamline will enable delivery of this low-current beam to SLAC's End Station A.  The beamline consists of (1) a long-pulse kicker and septum magnet to divert low-current bunches off the LCLS-II dump line, (2) a 250m long transfer line from the kicker in Sector 30 to the existing A-line, delivering beam to End Station A, (3) minor improvements in the existing End Station A infrastructure, and (4) an optional laser oscillator that augments the dark current with a well-defined, low- current beam at 46~MHz repetition rate within a $\sim$500 ns macro-pulse between LCLS-II/LCLS-II-HE primary bunches.  
LESA is being constructed in two stages: the first stage, S30XL, is nearly complete and will demonstrated the dark current extraction from the SRF linac while the 2nd stage, on which construction is just starting, will transport the extracted current to ESA.  S30XL will be commissioned in FY23 while the full LESA beamline is planned to be complete in FY24.

The presently envisioned program for LESA comprises dark sector physics, electron-nuclear scattering measurements for the neutrino program, and a test beam program (see \cite{S30XL_Science}).  

\section{PARAMETERS and POSSIBLE UPGRADES}

 The parameters of the LESA system are listed in Table 1. Three categories of parameters are listed: beam at the experiment, beam in the End Station A beamline at the spoiler/collimator system, and beam in the LCLS-II accelerator. The primary beamline is designed to meet the LDMX experimental requirements with an ultra-low current beam (i.e. sub-nA scale in the End Station) but has the capability of being upgraded to support ``low-current’’ (few uA) beams for Super-HPS type experiments. Table 1 lists parameters for the ultra-low current beam and then for the potential upgrade path. The ultra-low current beam has several applications beyond the LDMX experiment, including nuclear structure measurements motivated by the accelerator-based neutrino physics program and test beams.
 
For operation with ultra-low or low-current parameters, the LESA kicker extracts roughly 500 ns of bunches between the LCLS-II primary bunches spaced by 1.1 µs, as illustrated in Figure 3. In the case of the LDMX experiment, the desired electron current ranges between 100 fA and 150 pA, corresponding to from 1 to approximately 500 electrons per $\mu$s, or a maximum of 0.5 Watts of electron beam power at 4 GeV with a 55\% duty cycle. For the case of a low-current beam to support a Super-HPS type experiment, the beam current would be increased to an average of 1 $\mu$A which is still less than 4\% of the nominal maximum current in LCLS-II. In this case, the spoiler would not be used, and the maximum beam power would be less than 5 kW into End Station A. 

Finally, it is also possible for LESA to support a beam dump experiment where the LESA kicker timing would be shifted to extract the primary bunches that are not sent to the FEL undulators. In operation, the photon science experiments are expected to use roughly ½ of the total beam power delivered by the SRF linac and the excess power is expected to be roughly 900 MW-hr per year of electrons which would be sent to a beam dump in the BSY  \cite{LCLSIIHEbeamuseage}.  For the high-power Beam Dump experiments, these excess electron bunches could be deflected into the A-line by the LESA kicker and dumped upstream of End Station A. This upgrade would require installing a new 250 kW dump in the A-line, adding the appropriate shielding, and plugging the aperture passing from the A-line through the 6m shield wall into the End Station A enclosure.

Focusing on the ultra-low current parameters, the buckets to be extracted by the LESA kicker are filled at the RF gun. The LCLS-II RF gun specification is that dark current is less than 400 nA at 100 MeV. All of the LESA parameter sets stay below this current limit to exclude interference with LCLS-II. To ensure the performance required for LDMX and to enable higher currents in LESA (as for Super-HPS), a separate gun laser will likely be used. This laser would intentionally populate unused gun buckets at a sub-harmonic of the gun frequency. These bunches are well-separated from the primary beam bunches so they can be extracted by the LESA kicker downstream. The new gun laser would share the LCLS-II RF gun 46 MHz laser oscillator, but would have a separate amplifier, UV conversion, and transport, all of which operate at much lower average laser power than the LCLS-II systems.

\begin{table}
\begin{center}
\caption{LESA electron beam parameters for an ultra-low-current beam (baseline) as well a possible upgrade mode motivated by Super-HPS-style experiments; it is also possible to consider a high power mode where high power LCLS-II beam is dumped in the A-line rather than the BSY as described in the text. \\}
\label{tab:perf}
\begin{tabular}{l|c|c}
\hline\hline
ESA Beam Parameters& Ultra-low-current &	Low current  \\
&(baseline)&(upgrade)\\
\hline
Energy	& 2.0 - 8.0 GeV &
	2.0 - 8.0 GeV  \\
Bunch spacing&	21.5 ns &	5.4 ns\\
Bunch charge&	0.04 – 20 e-	&70,000 e- (10 fC)	\\
Macro pulse beam current	&0.1 – 150 pA	&2 $\mu$A\\
Macro pulse length / rep. rate & 1.1 $\mu$s / 929 kHz & 1.1 $\mu$s / 929 kHz \\
Duty cycle&	50\% &	50\% \\
Beam norm. emittance (rms)&	$~$300 $\mu$m& $<$ 1000 $\mu$m \\
Bunch energy spread&	$<$1\%& 	$<$1\%\\
IP spot size&	4 cm x 4 cm	&$<$250 $\mu$m including jitter\\
Max beam power&	0.5 W&	5 kW\\
\hline
	
ESA Spoiler Parameters	&&\\\hline		
Charge reduction&	0 – 99.99\%	&N/A	\\
Emittance increase	&1 - 1000x	&N/A	\\
Max beam power&	55 W&	N/A	\\
Spoiler thickness&	0 – 0.5 r.l.&	N/A	\\
\hline
			
Accelerator Parameters	&&\\\hline		
SRF Linac Energy	& 4.0 - 8.0 GeV &
	4.0 - 8.0 GeV  \\
Macro pulse beam current&	0 – 25 nA&	2 $\mu$A\\	
Beam norm. emittance (rms)	&1$\mu$m - 25 $\mu$m&	1$\mu$m \\
Beam admittance (edge)	&$<$50 nm	&$<$50 nm	\\
Bunch energy spread (FWHM)	&$<$2 \% &	$<$2 \%	\\
Bunch length (rms)&	$<$1 cm&	$<$1 cm	\\
Max beam power&	55 W&	5 kW\\
\hline\hline
\end{tabular}
\end{center}
\end{table}

\begin{figure}
\begin{center}
\includegraphics[width=0.95\hsize]{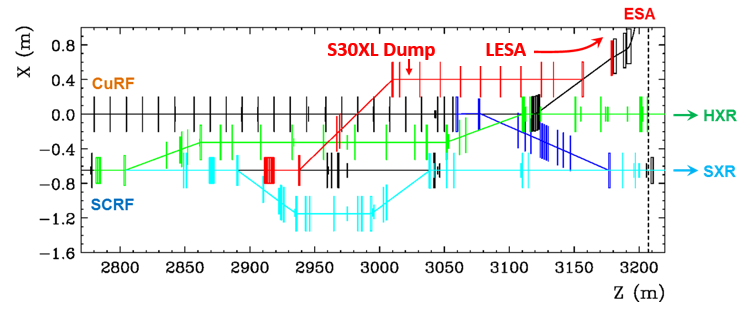}
\end{center}
\caption{Plan view of the S30XL/LESA beamline (red) placement at the end of the linac in the BSY along with the other LCLS and LCLS-II beamlines; all beamline components from the S30XL extraction kickers at 2920 m through the S30XL dump at 3030 m have been installed while work has started on fabrication of the downstream components of LESA.}
\label{fig:Fig4}
\end{figure}

\section{A-LINE AND END STATION A}
The A-line, which transports beams from the central part of the Beam Switch Yard (BSY) towards End Station A (ESA), was recently used to transport either the primary LCLS beam or a spoiled and collimated reduced number of electrons (secondary electron beam) from the LCLS beam into ESA. In 2013, the End Station Test Beam program (ESTB) \cite{ESTB} was established at SLAC to provide particle beams from the LCLS normal conducting Linac for detector R\&D experiments in ESA. The A-Line ESTB operated until 2019 with the full range of available LCLS-beams: electron beam energies between 2.5 GeV and 16.5 GeV and bunch charges between 20 pC and 280 pC.

In ESTB, the secondary electron beams were generated by steering selected LCLS bunches onto a target in the BSY. The resulting electron beam has a wide energy spread, which was then transported through the A-line into ESA. The thickness and material of the target were chosen appropriately to reduce the number of hadrons generated in the electron-target interaction to negligible levels. Additional spoilers are available to further diffuse the electron energy spread without generating a significant number of other particles. The A-Line bend magnets were set to the particle energy required by the experimenters in ESA and collimators are used to narrow the energy spectrum. Secondary particle beams have been delivered from 2 GeV up to the full-LCLS beam energy. 

LESA will use the same concept to control the electron beam current and spectrum.  A multi-collimator system in the A-line is used to control the number of electrons per pulse. A momentum slit reduces the accepted beam energy spread from about one percent to less than one part per million. Four-jaw collimators then reduce the geometrical spread of the accepted beam reaching experiments in End Station A.

The desired beam emittance for the LDMX experiment is large enough so that the beam can be defocused to a cross-section of roughly 4x4 cm. The desired beam emittance is many times (100-1000) that of the LCLS-II emittance (as well as the LCLS-II admittance, which is determined by the LCLS-II collimation system). This increase is accomplished using the ESA spoilers with a corresponding degradation of the beam current. Assuming an incoming, mono-energetic, 4-GeV beam, a simple 0.1 radiation length spoiler system increases the emittance to more than 300 $\mu$m, with more than 50\% of the current within 0.5\% of the incoming energy. In practice, LESA will have spoilers with different thicknesses which, combined with the downstream collimators, will control the beam emittance and current at the LDMX detector. The spoiler system has been specified for beam current up to 100 times higher than that needed at the experiment (i.e., 55 W) to allow options for precise control and shaping of the electron beam at the experiment.

\section{SUMMARY}
For an important class of light dark matter scenarios, electron fixed-target experiments have unparalleled sensitivity. The LESA beamline presents a unique, timely, and cost-effective opportunity to enable high-impact dark matter and dark force experiments. LESA can deliver a low-current, quasi-continuous electron beam into the existing End Station A (ESA) beamline by filling unused buckets from the LCLS-II linac, without impacting the LCLS-II program. LESA’s multi-GeV energy, high beam repetition rate and capability to host year-scale particle physics experiments offer a unique combination of advantages that make possible a wide range of world-class experiments.

\section{ACKNOWLEDGEMENTS}
The authors would like to thank John Jaros for helping make the connection on the LESA concept, the LDMX collaboration which has motivated the need for the facility, and the engineers and designers at SLAC that have developed the design. This work has been supported in part by DOE contract DE-AC02-76-SF00515.

\bibliographystyle{JHEP}
\bibliography{LESA}

\begin{thebibliography}{1}
\bibitem{S30XL_Science}
{A description of LESA and its science program can be found in SLAC-R-1147. The
  LESA beamline has been known as DASEL, S30XL, and YABBL.}
\end{thebibliography}
\end{document}